\documentclass[11pt]{article}
\PassOptionsToPackage{authoryear}{natbib}
\usepackage[preprint,nonatbib]{neurips_2024}
\usepackage{hyperref} 
\usepackage{authblk} 
\usepackage{booktabs}
\usepackage{graphicx}
\usepackage{subcaption}
\usepackage{amsfonts}
\usepackage{multicol}
\usepackage{tabularx}
\usepackage{enumitem,amsmath,bm,amssymb}
\usepackage{pifont,makecell}
\usepackage{tikz}
\usepackage{CJKutf8}

\usepackage{multirow}
\usepackage{cite}
\usepackage{comment}

\usepackage{svg}  

\usepackage{xcolor}
\usepackage{enumitem}

\title{End-to-End Training for Discrete Token LLM based TTS System}

\author{
  \bf Changfeng Gao, 
  \bf Yong Ren, 
  \bf Jun Yuan,
  \bf Ye Bai, 
  \bf Zhao You,
  \bf ShiDong Shang
}

\affil{
  Tencent Yuanbao, China
}
\date{}
\begin{document}

\maketitle

\begin{abstract}


Recent state-of-the-art (SOTA) text-to-speech (TTS) systems typically adopt a cascaded pipeline consisting of a speech tokenizer, an autoregressive large language model (LLM), and a diffusion based flow-matching (FM) model, with these components trained independently. In this paper, we propose a fully end-to-end (E2E) optimization framework that unifies the training of the speech tokenizer, LLM, FM model, and an additional reward model (RM).
Specifically, we first jointly optimize the tokenizer using multi-task objectives derived from reconstruction for FM, next-token prediction for LLM, and multi recognition task for RM. This joint training encourages the discrete speech token space to capture acoustically and semantically salient information that is better tailored to TTS. We then further optimize the LLM using downstream reconstruction and recognition by FM and RM, which reduces inference-time mismatch and steers the LLM toward more preferred generations.
Experimental results show that our E2E framework consistently outperforms cascaded baselines. On the Seed-TTS-Eval benchmark, our system achieves a word error rate (WER) of 0.78\% and 1.56\%, a new SOTA result with a 0.6B-parameter LLM and  0.5B-parameter FM model. These results validate that holistic E2E optimization is critical for improving discrete-token-based TTS systems with a much simpler training pipeline.
\end{abstract}

\section{Introduction}


In recent years, text-to-speech (TTS) synthesis has advanced rapidly, driven by the scaling of model capacity and training data. Among existing paradigms, the framework based on large language models (LLMs) operating over discrete acoustic tokens remains a dominant choice for recent state-of-the-art (SOTA) TTS systems, including Seed-TTS \cite{DBLP:journals/corr/abs-2406-02430}, CosyVoice3 \cite{du2025cosyvoice3}, and Qwen3-TTS \cite{hu2026qwen3}.
These systems typically consist of three modules: (1) a speech tokenizer that converts a speech signal into a sequence of discrete tokens, (2) an autoregressive LLM that predicts the speech-token sequence conditioned on input text, and (3) a flow-matching (FM) model that synthesizes acoustic features (or waveform) from the predicted tokens. In most existing systems, these components are designed and trained independently and then cascaded at inference time, which can introduce several limitations for TTS performance.


The first issue arises from the fact that the tokenizer must be pre-trained before the LLM and the FM model, and therefore largely determines the performance upper bound of the entire TTS system. Ideally, the discrete speech tokens should preserve (i) sufficient semantic information so that the LLM can reliably predict the token sequence from text, and (ii) sufficient acoustic detail so that the FM model can accurately reconstruct speech from the tokens.
To meet these requirements, a variety of speech tokenizers have been explored, including approaches that introduce quantization into automatic speech recognition (ASR) models \cite{ye2023asq,cosyvoice}, self-supervised learning (SSL) representations \cite{firered_tts}, and VAE-based speech autoencoders \cite{yu2025joyvoice,hu2026qwen3}, as well as hybrids that combine multiple paradigms. However, the objectives used to train these tokenizers (e.g., ASR accuracy, SSL representation quality, or autoencoding fidelity) are not necessarily aligned with the downstream autoregressive token prediction in the LLM or the denoising/reconstruction process in the FM model. This misalignment can lead to suboptimal end-to-end behavior. For instance, increasing the codebook count and size often improves token expressiveness and benefits ASR and FM reconstruction, but it also enlarges the prediction space and can substantially increase the learning difficulty for the LLM.


The second issue is that, due to the black-box nature of neural networks, it is difficult to precisely characterize what information a tokenizer actually encodes. Although tokens are often interpreted as being primarily semantic or acoustic—motivated by their intended downstream roles—the content of each token remains largely opaque and hard to disentangle. For example, prior studies \cite{gao2025diffro, ren2024emo, tseng2025codec2vec} have observed that so-called “semantic” tokens may also capture speaker attributes such as age, gender, or emotion, while tokens trained mainly for reconstruction can still be effective for ASR. These findings suggest that discrete speech tokens often contain substantial redundant, entangled, and potentially uncontrollable factors of variation, which can be detrimental to TTS by introducing ambiguity in token prediction and reducing controllability in synthesis.



Finally, because the modules in TTS systems are optimized independently, the FM model is typically trained with ground-truth speech tokens rather than tokens predicted by the LLM. As a result, errors and distributional shifts in the LLM outputs can propagate to the FM, leading to degraded synthesis quality at test time.
More fundamentally, the LLM is usually trained to maximize token prediction likelihood (i.e., reduce cross-entropy on discrete tokens) without explicitly accounting for the perceptual quality or expressiveness of the synthesized speech. This objective encourages the model to reproduce the dominant modes of the training distribution, often yielding safe and “average” generations. Such mode-seeking behavior is particularly detrimental for expressive TTS, where natural prosody and style require modeling diverse—and sometimes low-frequency—acoustic realizations beyond what is favored by token-level likelihood training.


Several recent studies have recognized these issues and attempted to address them through partial joint optimization of TTS components. For example, \cite{wang2026tadicodec, wang2026diffcodec} train the speech tokenizer with a diffusion-based speech reconstruction objective, allowing the FM model to directly shape the tokenizer toward representations that are optimal for reconstruction. JoyVoice \cite{yu2025joyvoice} jointly trains the LLM and FM model, showing that such coupling encourages the LLM to produce representations that are more “reconstruction-friendly” for the downstream generator. In addition, differentiable reward optimization (DiffRO) \cite{gao2025diffro}, adopted in CosyVoice, can be viewed as another form of joint training: it couples the LLM with the tokenizer backend to enable refinement via reward-driven learning.
Overall, these works consistently suggest that joint optimization can improve TTS performance. However, most existing approaches only co-train a subset of modules rather than optimizing the system holistically. Consequently, interface inconsistencies and train–test mismatches may still persist in the remaining independently trained components, limiting the potential gains from joint training.


In this paper, we propose a fully end-to-end (E2E) TTS framework that enables joint optimization of the speech tokenizer, the LLM, and the FM model.
First, we train the tokenizer with direct supervision from downstream components, including the FM-based speech reconstruction objective, the LLM token prediction objective, and a multi-task recognition reward model (RM). This design allows the tokenizer to learn representations that are explicitly shaped by the requirements of the overall TTS pipeline, rather than by proxy objectives (e.g., ASR/SSL) that may be misaligned with synthesis. We refer to the overall objective used in this stage as the first-order loss.
Besides the first-order loss, we sample token sequences from the LLM and feed them into the FM and RM, optimizing the system under the generated-token distribution to mitigate the train–test mismatch. Moreover, by using the Gumbel-Softmax, gradients can be propagated through the discrete sampling process, enabling direct optimization of the LLM with the reconstruction and recognition objectives. We refer to the corresponding objective as the second-order loss.


All the above issues can be alleviated through E2E optimization. The first-order objective enables the tokenizer to learn speech representations that are directly shaped by the needs of the TTS pipeline, rather than by potentially misaligned semantic/acoustic proxy tasks. The second-order objective further trains the LLM to produce reconstruction-friendly and recognition-consistent tokens under its own generation distribution, while also encouraging the FM model to be robust to error propagation from imperfect predicted tokens.
Experimental results validate the effectiveness of the proposed E2E training. Compared with separately trained pipelines, the E2E-optimized system achieves consistently better performance across all subtasks and yields higher overall TTS quality. On the Seed-TTS-Eval benchmark, our method attains word error rates (WER) of 0.78\% and 1.56\% using a 0.6B-parameter LLM and a 0.5B-parameter FM model, respectively. The reconstruction quality and the recognition accuracy of the FM and RM can also be improved.

\section{E2E TTS Training Framework}

\subsection{First Order Loss for TTS training}

\begin{figure}[t!]
    \centering
    \includegraphics[trim=3cm 14.8cm 3cm 5cm, clip, width=0.8\textwidth]{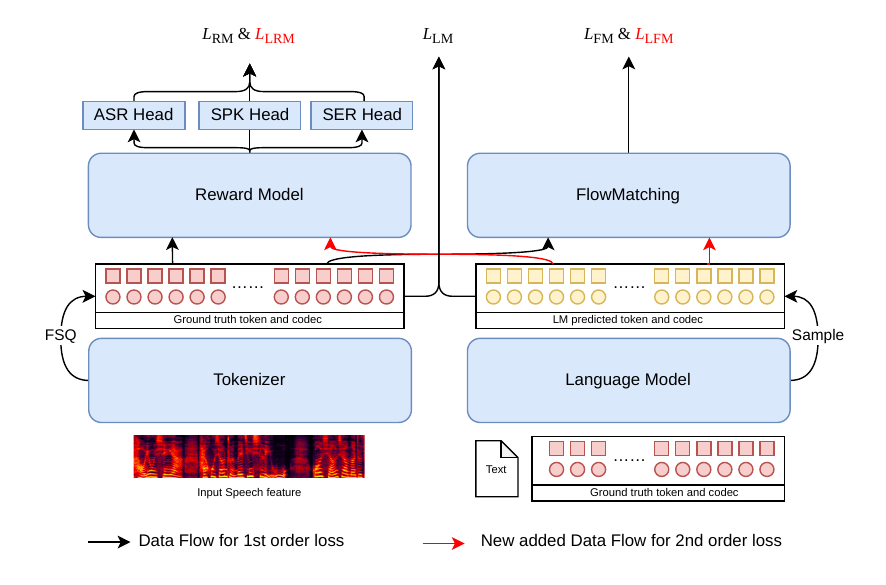}
    \caption{Illustrations of the E2E TTS framework. The LM, FM and RM can take both acoustic token and embedding as the input. And the FSQ codebook is shared between different modules and backpropagate the gradient in straight way. }
    \label{fig:e2etts}
\end{figure}

An illustration of the proposed E2E TTS framework is shown in Figure \ref{fig:e2etts}. The system consists of four components: a speech tokenizer, a multi-task recognition-based RM, a LLM, and a FM decoder. Given an input utterance $x_{1:T}$, the speech tokenizer encodes it into a discrete token sequence $c_{1:T}$ and a corresponding sequence of quantized representations $q_{1:T}$. Specifically, each token satisfies $c_i \in \{1,2,\dots,C\}$, and the associated quantized vector can be obtained by a codebook lookup:
\begin{equation}
q_i = \text{CodeBook}(c_i)
\end{equation}
For the quantization module, we adopt Finite Scalar Quantization (FSQ) \cite{DBLP:conf/iclr/MentzerMAT24} with a single codebook setup, which is commonly used in the TTS systems due to its simplicity and stable optimization.
\begin{equation}
    c_{1:T}, q_{1:T} = \text{Tokenizer}(x_{1:T})
\end{equation}
Then, we feed the representation sequence $q_{1:T}$ into the RM, FM, and LM to perform multi-task recognition, diffusion-based speech reconstruction, and next-token prediction, respectively. In the RM, we adopt ASR, speaker emotion recognition (SER), and speaker identification (SPK) as downstream tasks. These tasks share a Transformer-based audio encoder with different head and are jointly trained using CTC loss, cross-entropy (CE) loss, and a cosine similarity loss, as follows:
\begin{equation}
\begin{aligned}
&h_{1:T} = \text{Encoder}_{\text{RM}}(q_{1:T}) \\
&L_{\text{ASR}} = \text{CTC}\bigl(y_{1:N}, \text{Head}_{\text{ASR}}(h_{1:T})\bigr) \\
&L_{\text{SER}} = \text{CrossEntropy}\bigl(e, \text{Head}_{\text{SER}}(h_{1:T})\bigr) \\
&L_{\text{SPK}} = \text{CosSimilarity}\bigl(\text{x-Vector}(x_{1:T}), \text{Head}_{\text{SPK}}(h_{1:T})\bigr)
\end{aligned}
\end{equation}
where $y_{1:N}$ and $e$ denote the speech transcription and the emotion label, respectively, and the x-Vector model is a pretrained speaker-embedding extractor. Finally, we sum the ASR, SER, and SPK losses to obtain the RM loss:

\begin{equation}
    L_{\text{RM}} = L_{\text{ASR}} + L_{\text{SER}} + L_{\text{SPK}}
\end{equation}\label{eq:loss_rm}

Similar to other TTS system, the FM decoder is also trained to reconstruct $x_{1:T}$ conditioned on the discrete representation, but directly refuse the ${q_{1:T}}$ rather than a new embedding layer. It learns to predict a velocity field that transforms noisy samples back to the original speech as:
\begin{equation}
{x}_{1:T} = \mu{x}_{1:T} + (1-\mu)n_{1:T}, \text{ where } n_{i} \sim \mathcal{N}(\mathbf{0}, \mathbf{I}) \text{ and } \mu \sim U(0, 1)
\end{equation}
The training object of the FM is to predict the true velocity:
\begin{equation}
    L_{\text{FM}} = \lVert \text{FM}(q_{1:T}, \mu) - (x_{1:T} - n_{1:T})\rVert
\end{equation}\label{loss_fm}
Finally, we jointly train the LLM with the speech tokenizer to learn discrete representations that are amenable to autoregressive prediction. The LM also takes $q_{1:T}$ as input and predicts the distribution of the next token conditioned on the preceding history and the text:
\begin{equation}
\begin{aligned}
    &p(c_{t}|c_{1:t-1}), h_i^{\text{LM}} = \text{LM}(q_{1:t-1}, y_{1:N}) \\
    &L_{\text{LM}} = \sum_{t=1}^T\text{CrossEntropy}(c_{t}, p(c_{t}|c_{1:t-1}))
\end{aligned}
\end{equation}
where $h_{\text{LM}}$ denotes the final hidden states of the LM. However, because the LM and the tokenizer are jointly trained, the target label $c_{i+1}$ changes throughout training, which can destabilize optimization. To address this, once the tokenizer becomes trainable, we replace the logits with the similarity between $h_i^{\text{LM}}$ and the discrete codebook entries, and we stop the gradients flowing into the codebook.
\begin{equation}
\begin{aligned}
    p(c_t = K|c_{1:t-1}) = \frac{\exp(\text{sim}(h_t^{\text{LM}}, \text{CodeBook}(K)))}{\sum_{k=1}^C\exp(\text{sim}(h_t^{\text{LM}}, \text{CodeBook}(k)))}
\end{aligned}
\end{equation}

Since all loss functions proposed in this subsection take the ground-truth speech tokens as inputs, their gradients can be backpropagated into the speech tokenizer. We refer to them as the first-order loss, denoted by $L_1$:
\begin{equation}\label{eq:l1_loss}
    L_{1} = \alpha L_{\text{LM}} + \beta L_{\text{RM}} + \gamma L_{\text{FM}}.
\end{equation}
The loss $L_1$ is commonly used in most TTS systems. In our E2E-TTS framework, it can also be used to optimize the speech tokenizer.

\subsection{Second Order Loss for TTS Reinforcement}


In addition to the ground-truth speech tokens, both the FM and RM can also take the tokens predicted by the LLM as inputs. This strategy mitigates the mismatch between training and inference in the FM pipeline. Moreover, when gradients from the FM and RM are permitted to backpropagate into the LLM, the model is explicitly encouraged to generate speech tokens that are not only more accurate but also better aligned with the requirements of reconstruction.

Specifically, we replace the ground-truth tokens $q_{1:T}$ with either the Gumbel–Softmax hard samples or the hidden states of the LLM, denoted as $h^{\text{LM}}_{1:T}$:
\begin{equation}
    q'_t = \text{CodeBook}\bigl(\text{GumbelSoftmax}(p(c_t \mid c_{1:t-1}))\bigr)
    \quad\text{or}\quad
    h^{\text{LM}}_t.
\end{equation}

Accordingly, the reconstruction and flow-matching losses computed using LLM-predicted tokens are defined as $L_{\text{LRM}}$ and $L_{\text{LFM}}$, respectively. The combination of these terms constitutes the second-order loss $L_2$:
\begin{equation}
    L_2 = L_{\text{LRM}} + L_{\text{LFM}}.
\end{equation}

By incorporating $L_2$, the LLM outputs are optimized under the guidance of both the RM and FM objectives. This process can be interpreted as a form of reinforcement-style alignment, where the model improves along multiple dimensions, including pronunciation accuracy  emotional expressiveness and speaker similarity (SS).

\subsection{Training Pipeline} \label{sec:training}

\begin{figure}[t!]
    \centering
    \includegraphics[trim=0cm 21.5cm 0cm 1cm, clip,width=1.0\textwidth]{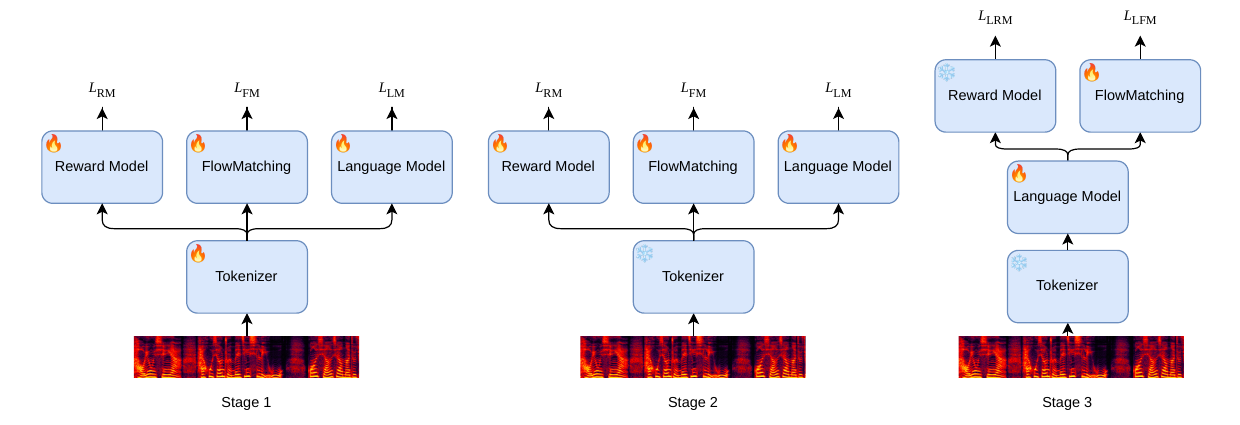}
    \caption{Three stage training for the E2E TTS.}
    \label{fig:train_stage}
\end{figure}

We split the E2E training pipeline into three stages.
At the first stage, all parameters of the E2E system are trainable, and the $L_1$ loss is used. This stage aims to train the speech tokenizer with the RM, FM and the LM downstream tasks with $L_1$. After Stage 1, we further freeze the speech tokenizer and then train the RM, FM and LM independently. And in this stage, we can use different training dataset for the RM, FM and LM according to their requirenmets. For example, we can add more noisy and low sampling rate speech for the RM training and choose more clean and broadband for FM and RM. And at final stage 3, we further freeze the RM and train the LM and the FM with the $L_2$.

\subsection{Analysis for the E2E training}\label{sec:analysis}

In LLM-based TTS systems, the speech tokenizer serves as a critical component that fundamentally determines the upper bound of system performance. Prior studies have shown that an effective speech tokenizer should encode both acoustic and semantic information, and various training strategies have been explored to capture these properties. However, few works have analyzed the quality of speech tokens from a theoretical and mathematical perspective.

In this work, we formulate speech tokenizer training as a source coding problem, allowing the information captured by each speech token to be quantified using information entropy:
\begin{equation}
    H(X_n) = -\sum_{k=1}^{C} p(X_n = k) \log_2 p(X_n = k).
\end{equation}

To maximize \( H(X_n) \), we consider the relationship between the input speech \( Y \) and the discrete tokens \( X \):
\begin{equation}
    H(Y_n) = H(X_n) + H(Y_n \mid X_n) - H(X_n \mid Y_n).
\end{equation}

Since the speech tokenizer performs a deterministic mapping, we have \( H(X_n \mid Y_n) = 0 \). Consequently, maximizing \( H(X_n) \) is equivalent to minimizing the conditional entropy \( H(Y_n \mid X_n) \). This term can be interpreted as the reconstruction uncertainty and is naturally modeled through a speech reconstruction objective. In our framework, \( H(Y_n \mid X_n) \) is minimized via the flow matching loss \( L_{\text{FM}} \), which provides a more stable and robust optimization procedure by leveraging ordinary differential equations (ODEs).

Furthermore, LLM-based TTS systems generate speech tokens in an autoregressive manner, implying that the token sequence should be modeled as a source with memory rather than an i.i.d. process. For analytical tractability, we adopt a first-order Markov source model, for which the lossless compression limit is governed by the entropy rate:
\begin{equation}
    \tilde{H}(X) = H(X_{n+1} \mid X_n) = H(X_n) - I(X_{n+1}; X_n).
\end{equation}

This formulation indicates that historical tokens must retain sufficient mutual information \( I(X_{n+1}; X_n) \) to reduce the uncertainty of future tokens. While prior works often rely on auxiliary ASR tasks to extract temporal semantics—effectively converting \( X_n \) into a latent Markovian representation \( T_n \) and modeling \( I(T_{n+1}; T_n, X_{1:N}) \), our E2E TTS framework optimizes \( I(X_{n+1}; X_n) \) directly using the language modeling loss \( L_{\text{LM}} \), without requiring intermediate hidden representations.

In summary, a well-designed speech tokenizer should achieve both high token entropy \( H(X_n) \) and strong temporal dependency modeling \( I(X_{n+1}; X_n) \). This implies that the codebook should be fully utilized, while past tokens should provide sufficient contextual information. These objectives are inherently consistent with existing practices that aim to encode both acoustic fidelity and semantic richness within speech tokens.

\section{Experiment}
\subsection{Experiments Setup}
\subsubsection{DataSets}

We conduct experiments on a large-scale in-house TTS dataset comprising approximately 100,000 hours of speech in both Chinese and English, with a language ratio of 4:1. All speech data are collected from publicly accessible websites and transcribed using Whisper-Large-V3 and FireRedASR. In addition, DNSMOS is employed to estimate the perceptual quality of each utterance. To reduce the impact of background noise and automatic transcription errors, we apply strict data filtering criteria: utterances with a DNSMOS score below 3.3 or a transcription difference rate exceeding 5\% are removed from the training set.

\subsubsection{Models}

For the speech tokenizer and the reward model, we construct the speech tokenizer and RM by inserting an FSQ-based quantization module into an existing CTC-based ASR system equipped with a Conformer encoder. The quantizer is configured with a low-rank dimensionality of 8, where each dimension contains 3 codes, resulting in a total codebook size of 6561. To support additional downstream tasks, we introduce two attention pooling heads for SER and SPK. The total parameter counts of the tokenizer and the RM are 0.6B and 0.7B, respectively.

The LM and the FM of the E2E-TTS follows mainstream end-to-end TTS pipelines. For the LM, we adopt Qwen3-0.6B and extend its vocabulary with 6561 speech tokens and 4 special control tokens, denoting the beginning and end of text and speech sequences, respectively. The speech token embeddings are tied to the FSQ codebook via a lightweight linear adaptation layer. For the FM, we employ a Diffusion Transformer (DiT) architecture following CosyVoice3. The LM and FM contain 0.6B and 0.5B parameters, respectively. Finally, a HiFiGAN vocoder is used to synthesize waveforms from predicted Mel spectrograms.

\subsubsection{Training Detail}


We train the end-to-end TTS system in three stages, as described in Section \ref{sec:training}, using 64 NVIDIA H20 GPUs.

Stage 1. We adopt the AdamW optimizer with a linear warmup and cosine annealing schedule. The warmup phase consists of 25k steps, and the peak learning rate is set to \(1\times10^{-4}\). The loss weights for \(L_{\text{LM}}\), \(L_{\text{FM}}\), and \(L_{\text{RM}}\) are set to 0.1, 1.0, and 1.0, respectively.

Stage 2. The learning rate is reduced to \(1\times10^{-5}\), while maintaining a 25k-step warmup. In this stage, we improve the robustness of the reward model by balancing the training data: emotion labels predicted by Emo2Vec-Large are incorporated, and additional noisy speech samples are introduced to enhance ASR-related supervision.

Stage 3. We introduce the second-order loss \(L_2\) for end-to-end TTS optimization. Within \(L_{\text{LRM}}\), the weights for the speaker loss \(L_{\text{SPK}}\) and the ASR loss \(L_{\text{ASR}}\) are set to 0.1 and 1.0, respectively. The emotion loss \(L_{\text{SER}}\) is applied only when emotion conditioning is required.

\subsection{Experiments Results}

\subsubsection{Main Results for Zero-Shot TTS}

\begin{table*}[t!]
	\centering
	\setlength\tabcolsep{4.5pt}
	\scalebox{1.0}{
		\begin{tabular}{lcccccc}
			\toprule
			\multirow{2}{*}{\textbf{Model}} & \multicolumn{2}{c}{\textbf{\emph{test-zh}}} & \multicolumn{2}{c}{\textbf{\emph{test-en}}} & \multicolumn{2}{c}{\textbf{\emph{test-hard}}} \\
			\cmidrule(r){2-3} \cmidrule(r){4-5} \cmidrule(r){6-7}
			& \textbf{CER (\%)~$\downarrow$} & \multicolumn{1}{c}{\textbf{SS~$\uparrow$}} & \textbf{WER (\%)~$\downarrow$} & \multicolumn{1}{c}{\textbf{SS~$\uparrow$}} & \textbf{CER (\%)~$\downarrow$} & \multicolumn{1}{c}{\textbf{SS~$\uparrow$}}  \\
			\midrule
			\textbf{Human} & 1.26 & 0.755 & 2.14 & 0.734  & - & -- \\
			\midrule
                \multicolumn{7}{c}{\textbf{Non-autoregressive Models}} \\
                \midrule
			\textbf{MaskGCT}~\cite{DBLP:journals/corr/abs-2409-00750} & 2.27 & 0.774 & 2.62 & 0.714  & 10.27 & 0.748 \\
			\textbf{F5-TTS}~\cite{DBLP:journals/corr/abs-2410-06885} & 1.56 & 0.741 & 1.83 & 0.647  & 8.67 & 0.713 \\
            \textbf{F5R-TTS}~\cite{sun2025f5r} & 1.37 & 0.754 & -- & --  & 8.79 & 0.718 \\
            \textbf{ZipVoice}\cite{zhu2025zipvoice} & 1.40 & 0.751 & 1.70 & 0.697  & -- & -- \\
            \textbf{OmniVoice}\cite{zhu2026omnivoice} & 0.84 & 0.777 & 1.60 & 0.741 & -- & --  \\
            \midrule
            \multicolumn{7}{c}{\textbf{Autoregressive Models}} \\
            \midrule
            \textbf{Seed-TTS}~\cite{DBLP:journals/corr/abs-2406-02430} & {1.12} & \textbf{0.796} & 2.25 & \textbf{0.762}  & 7.59 & \textbf{0.776} \\
            \textbf{Spark TTS}\cite{wang2025spark} & 1.20 & 0.672 & {1.98} & 0.584  & -- & -- \\
            \textbf{CosyVoice3-1.5B}\cite{du2025cosyvoice3} & {1.12} & 0.781& 2.21 & 0.720& 5.83 & 0.758 \\
            \textbf{Qwen3-TTS-0.6B}\cite{hu2026qwen3} & 1.18 & -- & 1.64 & -- & -- & -- \\
            \textbf{JoyVoice}\cite{yu2025joyvoice} & 0.97 & 0.786 & 1.69 & 0.736 & 5.55 & 0.746 \\
            \midrule
            \textbf{E2E-TTS-Stage1} & 1.16 & 0.775 & 2.09 & 0.690 & 7.68 & 0.752 \\
            \textbf{E2E-TTS-Stage2} & 0.86 & 0.775 & 1.72 & 0.696 & 7.21 & 0.752 \\
            \textbf{E2E-TTS-Stage3} & \textbf{0.78} & 0.781 & 1.56 & 0.705 & 6.61 & 0.759 \\
            w/o E2E-training & 0.87 & 0.760 & 1.89 & 0.682 & 7.35 & 0.745  \\
			\bottomrule
	\end{tabular}}
	\caption{Zero-shot TTS performance comparison between E2E-TTS and other TTS systems on the SEED test sets in terms of content consistency (WER/CER) and speaker similarity (SS).}
	\label{tab:tts-seed-test}
\end{table*}



We evaluate our end-to-end TTS system on the SEED-TTS test set, and the results are summarized in Table~\ref{tab:tts-seed-test}. As shown, our system achieves comparable or better performance to recent state-of-the-art TTS models in terms of both WER/CER and SS.
The results also highlight the importance of each training stage within the proposed end-to-end framework, particularly for pronunciation accuracy. In Stage~1, both the input and output of the LM remain variable, leading to relatively unstable LM outputs. Nevertheless, this stage effectively guides the tokenizer to produce token sequences with reduced entropy. Between Stages~2 and 3, a notable improvement in speaker similarity is observed, attributed to the joint training of the LM and FM modules, allowing them to better adapt to one another. Furthermore, the RM helps refine the LM’s preferences, mitigating overfitting during later stages.

To further examine the impact of end-to-end optimization, we conduct an ablation study in which the tokenizer is pretrained independently, after which the LM and FM are trained from scratch without joint optimization, following conventional TTS pipelines. Even with identical training data, this setup yields inferior performance on both WER and SS compared to the full end-to-end training strategy, underscoring the critical role of joint optimization. Future work will investigate the effects of end-to-end training on individual components—including the FM, RM and the speech tokenizer.

\subsubsection{Reconstruction Ability for FM}

\begin{table*}[!t]
	\centering
	\setlength\tabcolsep{4.5pt}
	\scalebox{1.0}{
		\begin{tabular}{lcccccc}
			\toprule
			\multirow{2}{*}{\textbf{Model}} & \multicolumn{2}{c}{\textbf{\emph{test-zh}}} & \multicolumn{2}{c}{\textbf{\emph{test-en}}} & \multicolumn{2}{c}{\textbf{\emph{CV3-Subject}}} \\
			\cmidrule(r){2-3} \cmidrule(r){4-5} \cmidrule(r){6-7}
			& \textbf{CER (\%)~$\downarrow$} & \multicolumn{1}{c}{\textbf{SS~$\uparrow$}} & \textbf{WER (\%)~$\downarrow$} & \multicolumn{1}{c}{\textbf{SS~$\uparrow$}} & \textbf{WER (\%)~$\downarrow$} & \multicolumn{1}{c}{\textbf{SS~$\uparrow$}}  \\
			\midrule
			$S^3$\textbf{-Tokenizer-FSQ}\cite{du2024cosyvoice2} & 3.31 & 0.787 & 4.09 & 0.709 & 11.50 & 0.761  \\
            \textbf{E2E-TTS-Stage1} & 3.36 & 0.814 & 3.42 & 0.688 & 11.71 & 0.762 \\
            \textbf{E2E-TTS-Stage2} & 3.05 & 0.825 & 3.26 & 0.704 & 11.67 & 0.781 \\
            \textbf{E2E-TTS-Stage3} & 2.93 & 0.826 & 3.08 & 0.702 & 11.56 & 0.780 \\
            \midrule
            w/o $L_\text{{LM}}$ & 3.12 & 0.812 & 3.45 & 0.691 & 11.56 & 0.774 \\
            w/o $L_\text{{LM}}$ and $L_\text{{FM}}$ & 3.68 & 0.799 & 4.11 & 0.662 & 12.30 & 0.768 \\
			\bottomrule
	\end{tabular}}
	\caption{Speech reconstruction performance comparison on the SEED test sets and CV3-Subject test set in terms of content consistency (WER/CER) and speaker similarity (SS). }
	\label{tab:result-fm}
\end{table*}



We evaluate the speech reconstruction performance of the FM module across different training stages on the SEED-TTS test set. Since the SEED-TTS utterances are predominantly read-style speech, we additionally assess performance on the CV3-EVAL subject test set, which comprises 178 speech samples spanning diverse genders, ages, emotions, speakers, and speaking styles.  
As shown in the results, a substantial improvement in both WER and SS occurs between Stage 1 and Stage 2. This trend aligns with observations reported in other studies involving joint training of tokenizers and FM modules \cite{wang2026tadicodec, wang2026diffcodec}. However, in Stage 3, improvements are observed only in WER, while SS remains largely unchanged. This behavior is expected, as Stage 3 adapts the FM to the LM outputs, which primarily enhance pronunciation accuracy rather than speaker similarity.  

We further investigate the roles of $L_{\text{FM}}$ and $L_{\text{LM}}$ in Eq.~\eqref{eq:l1_loss}. In Stage 1, we train the model without $L_{\text{FM}}$ or $L_{\text{LM}}$, and subsequently train the FM independently in Stage 2. The results demonstrate that both losses are essential, as they supply acoustic and semantic information to the speech tokenizer. Although besides the sematic information learned from ASR task, $L_{\text{RM}}$ can also contribute acoustic information via SER and SPK tasks, $L_{\text{FM}}$ proves to be particularly critical for the quality of speech reconstruction.

\subsubsection{Recognition Ability for RM}

\begin{table*}[htb]
	\centering
	\setlength\tabcolsep{4.5pt}
	\scalebox{1.0}{
		\begin{tabular}{lcccccc}
			\toprule
			\multirow{2}{*}{\textbf{Model}} & \multicolumn{3}{c}{\textbf{\emph{ASR WER (\%)}}} & \multicolumn{2}{c}{\textbf{\emph{SER WA (\%)}}} \\
			\cmidrule(r){2-4} \cmidrule(r){5-6} 
			& CMV-zh & CMV-en & LS-clean & IEMOCAP & MELD \\
			\midrule
            \textbf{Whisper-large-V3}\cite{systran_fwhisper_large_v3} & 12.40 & 9.66 & 2.56 & -- & -- \\
            \textbf{Emo2Vec-large} & -- & -- & -- & 67.3 & 57.4 \\
            \textbf{SenseVoice-small}\cite{funaduiollm} & 10.78 & 14.71 & 3.15 & 65.7 & 57.8 \\
            $S^3$\textbf{-Tokenizer-FSQ}\cite{du2024cosyvoice2} & 7.27 & 10.67 & -- & -- & -- \\
            \textbf{Qwen-TTS-Tokenizer}\cite{hu2026qwen3} & 14.99 & 10.40 & -- & -- & -- \\
            \textbf{E2E-TTS-Stage1} & 6.85 & 14.62 & 2.27 & -- & -- \\
            \textbf{E2E-TTS-Stage2} & 6.50 & 14.53 & 2.22 & 60.8 & 55.6 \\
            \midrule
            w/o $L_\text{{LM}}$ & 7.16 & 14.77 & 2.23 & 60.2 & 54.1 \\
            w/o $L_\text{{LM}}$ and $L_\text{{FM}}$ & 7.69 & 16.53 & 2.96 & 59.5 & 55.7 \\
			\bottomrule
	\end{tabular}}
	\caption{ASR and SER performance for different models. }
	\label{tab:result-asr-ser}
\end{table*}


The ASR and SER capabilities of the RM are presented in Table~\ref{tab:result-asr-ser}. For ASR evaluation, we report the WER on the LibriSpeech test-clean set and the CommonVoice test set. For SER, we measure the Weighted Accuracy (WA) on the IEMOCAP and MELD datasets. Despite operating on discretized inputs, the RM demonstrates robust performance on these downstream recognition tasks, outperforming the $S^3$-tokenizer and Qwen-TTS-tokenizer on the Chinese ASR task. Notably, performance gains are observed between Stage 1 and Stage 2, indicating that standalone fine-tuning is beneficial following joint training.
Regarding the impact of $L_{\text{FM}}$ and $L_{\text{LM}}$, we observe that both losses contribute positively to the RM’s performance. Their effect is analogous to self-supervised learning paradigms, such as TERA~\cite{liu2021tera} and VQ-Wav2Vec~\cite{baevski2019vqwav2vec}, where auxiliary objectives enhance representation learning.

\subsubsection{Statistical Characteristic Analysis for Speech Token}

\begin{figure}[!t]
    \centering
    \begin{subfigure}{0.28\linewidth}
        \includegraphics[width=\linewidth]{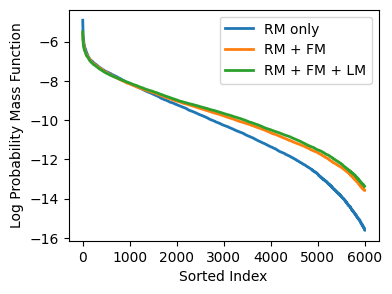}
        \caption{Distribution for $X_n$}
        \label{fig:left}
    \end{subfigure}%
    \hfill
    \begin{subfigure}{0.70\linewidth}
        \includegraphics[width=\linewidth]{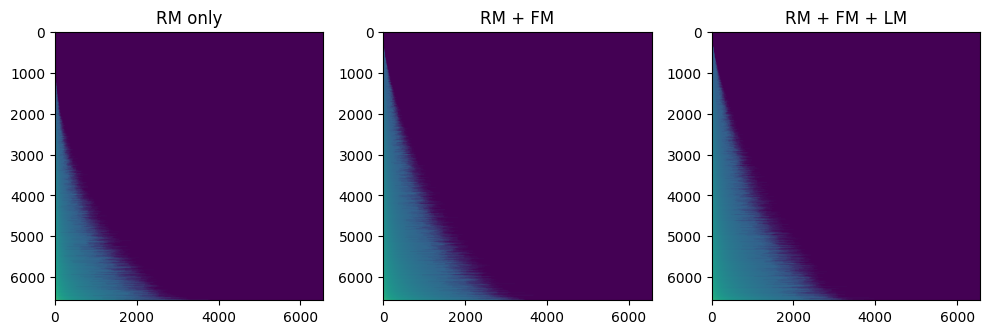}
        \caption{Joint Distribution for $X_{n+1}X_n$}
        \label{fig:right}
    \end{subfigure}
    \caption{Distributions for speech token trained with different loss functions on the LibriTTS training set.}
    \label{fig:two_subfigs}
\end{figure}

\begin{table*}[!t]
	\centering
	\setlength\tabcolsep{4.5pt}
	\scalebox{1.0}{
		\begin{tabular}{lcccccc}
			\toprule
			Tokenizer & $H(X_n)$ &  $I(X_{n+1};X_n)$ \\
			\midrule
            RM only & 10.905 & 2.50 \\
            RM + FM & 11.254 & 2.53  \\
            RM + FM + LM & 11.363 & 2.55 \\
			\bottomrule
	\end{tabular}}
	\caption{Statistical characteristic for the speech token trained with different loss functions on the LibriTTS training set. }
	\label{tab:result-entropy}
\end{table*}

Finally, we analyze the statistical characteristics of speech tokens on the LibriTTS training set. The distributions are visualized in Figures \ref{fig:left} and \ref{fig:right}, where tokens have been sorted by frequency to enhance clarity. We can find that with the help of the FSQ, nearly 100\% utilization of the codebook can be achieved for all tokenizers, but without $L_{\text{LM}}$ and $L_{\text{FM}}$, the token indices are heavily concentrated within a narrow range, exhibiting a severe long-tail distribution. While $L_{\text{FM}}$ effectively flattens the uni-gram distribution, $L_{\text{LM}}$ further smooths the bi-gram distribution, indicating that the tokenizer successfully captures richer acoustic information as the flatten distribution always has higher information entropy.

We further compute $H(X_n)$ and $I(X_{n+1}; X_n)$ as defined in Section \ref{sec:analysis}. As shown in Table \ref{tab:result-entropy}, the E2E-trained tokens exhibit higher values for both metrics. This suggests that the speech tokenizer utilizes the codebook more efficiently and that historical tokens carry more predictive information. Consequently, the FM and LM are better equipped to reconstruct high-quality speech and accurately predict subsequent tokens, which are the fundamental objectives of a TTS system.

\section{Conclusion}
In this paper, we propose an E2E training framework for LLM-based TTS systems. Within this framework, the speech tokenizer, the LM, the FM, and an auxiliary RM are jointly optimized, allowing the tokenizer to directly encode speech according to the LM and FM requirements. And the FM and RM can directly guide the LM predict the token according to the pronunciation accuracy and speech quality. Furthermore, we design a three-stage training pipeline to make the training process stable and effective. Experimental results demonstrate that, benefiting from E2E optimization, the proposed TTS system achieves WERs of 0.78\% and 1.56\% in the Seed-TTS zh and en test sets, respectively, which is comparable to the latest SOTA TTS systems with a much simplified training method.

\bibliographystyle{unsrt}
\bibliography{main}

\appendix

\end{document}